\documentclass[referee]{chjaa}          % referee version: for submission
\usepackage{natbib}
\usepackage{txfonts}
\usepackage[dvips]{graphicx}
\begin{document}

   \title{The debeamed luminosity, sychrotron peak frequency and black hole
mass of BL Lac objects
%\footnotetext{$*$ Supported by the National Natural Science Foundation of China.}
}
%   \subtitle{I. Place Your Subtitle Here}

   \volnopage{Vol.0 (200x) No.0, 000--000}      %%preserved for Editor. DOn't remove!
   \setcounter{page}{1}           %%starting page, preserved for Editor. DOn't remove!

   \author{Zhongzu Wu
      \inst{1,2,3}\mailto{}
%% Please move "\mailto{}" to the corresponding author of the paper.
   \and Minfeng Gu
      \inst{1,2}
   \and D. R. Jiang
      \inst{1,2}
      }

   \institute{Shanghai Astronomical Observatory, Chinese
Academy of Sciences, Shanghai 200030, China\\
             \email{zzwu@shao.ac.cn}
%% offprint requests will be sent. Note to pair \mailto{} with \email{}
        \and
             Joint Institute for Galaxy and Cosmology (JOINGC) of SHAO and
USTC\\
        \and
             Graduate School of the Chinese Academy of Sciences, Beijing
100039\\
          }

 %  \date{Received~~2001 month day; accepted~~2001~~month day}

   \abstract{
We estimate the intrinsic luminosities and synchrotron peak
frequency using the derived Doppler factor for a sample of 170 BL
Lac objects, of which the synchrotron peak frequency are derived by
fitting the SED constructed with the collected multi-band data from
literatures. We find that the debeamed radio and optical core
luminosities follow the same correlation found for FR I radio
galaxies, which is in support of the unification of the BL Lac
objects and the FR I galaxies based on orientation. For the debeamed
luminosity at synchrotron peak frequency, we find a significant
positive correlation between the luminosity and intrinsic
synchrotron peak frequency. This implies that the more powerful
sources may have the majority of jet emission at higher frequency.
At synchrotron peak frequency, the intrinsic luminosity and black
hole mass show strong positive correlation, while mild correlation
is found in the case of jet power, indicating that the more powerful
sources may have heavier black hole. %Moreover, the debeamed
%luminosities well follow the fundamental plane found for AGNs and
%X-ray binaries.
   }

   \authorrunning{Zhongzu, Wu, Minfeng Gu  \&D. R. Jiang }            %author_head in even pages
   \titlerunning{Wu et al. }  % title_head in odd pages

   \maketitle
%% The author head (on even pages) and the title head (on odd pages) will be
%% automatically extracted from \author{} and \title{}. Whenever the title is too long,
%% you will be asked to supply a shorter one by inserting either \authorrunning{} or
%% \titlerunning{} before \maketitle. Anyway, you can specify your own heads in advance.
%%
%%
%% Note: In the following text body of your manuscript, please note several differences from
%%       other major journals:
%% (1) \subsection{Please Capitalize the First Letter of Each Notional Word in Subsection Title}
%% (2) Please Capitalize the First Letter of Each Notional Word in all tables' captions

%
%________________________________________________ sections below
%
\section{Introduction}           %% first-level sections will be auto-capitalized
BL Lac objects are one of the most extreme classes of active
galactic nuclei (AGN) with no emission lines or less of them, but
they have a strong highly variable and polarized non-thermal
continuum emission ranging from radio to $\gamma$-ray bands.
Traditionally, they have been divided into radio-selected and
X-ray-selected BL Lac objects (RBLs and XBLs, respectively) based
on the surveys they were primarily discovered. In recent years,
the dichotomy has been replaced by a more physically meaningful
classification based upon the overall spectral energy distribution
(SED) of the object, namely, low frequency peaked BL Lac objects
(LBL), intermediate objects (IBL) and high frequency peaked BL Lac
objects (HBL) \citep{padovani95}. Generally, RBLs tend to be LBLs,
and XBLs tend to be HBLs %the LBLs and HBLs are observed at
%different viewing angles
(see review by \cite{urry95}). The synchrotron peak frequency of
RBLs is usually in the radio/IR band, while UV/X-ray band for XBLs
\citep{Giommi95}. The recent studies on the SED of large samples of
BL Lac objects shown that BL Lac objects most likely form one class
with a continuous distribution of synchrotron emission peak
energies, while RBLs and XBLs represent the opposite ends of the
continuum \citep{nieppola2006}.

%\cite{sambr96}
%showed that the typical multi-wave-band SED of LBLs and HBLs can
%not be accounted for in terms of a sole change in viewing angle
%but requires a systematic change of intrinsic physical parameters.
\cite{fossati98} and \cite{ghisel98} have proposed the so-called
`blazar sequence' with plots of various powers vs. the synchrotron
peak frequency $\nu_{\rm peak}$ for a sample of blazars containing
flat-spectrum radio quasars (FSRQ) and BL Lac objects, of which an
anti-correlation was apparent, with the most powerful sources having
relatively small synchrotron peak frequencies and the least powerful
ones having the highest $\nu_{\rm peak}$. The theoretical
interpretation to these anti-correlations was proposed by
\cite{ghisel98}: the more powerful sources suffered a larger
probability of losing energy and therefore are subjected to more
cooling and therefore translates into a lower value of $\nu_{\rm
peak}$. However, the recent studies have largely changed this
scenario. Recently, \cite{nieppola2006} proposed that there are
anti-correlations between the luminosity at 5 GHz, 37 GHz, and 5500
$\dot{\rm A}$ and $\nu_{\rm peak}$ for a large sample of BL Lac
objects, while based on all the correlations with $\nu_{\rm peak}$,
they concluded that the blazar sequence scenario is not valid.
Through estimating the Doppler factor for a sample of 170 BL Lac
objects selected from \cite{nieppola2006}, \cite{wu06} found a
significant anti-correlation between the total 408 MHz luminosity,
which is supposed to be intrinsic, and the intrinsic $\nu_{\rm
peak}$, with a large scatter which however undermines the `blazar
sequence'.
%, as a whole, LBLs are found to be more powerful than HBLs.
%Moreover, the authors
%claimed that HBLs have smaller Doppler factor and larger viewing
%angle than LBLs in a large BL Lac objects sample.
Moreover, the evidence of low-power LBLs has been recently
discovered (Padonavi et al. 2003; Caccianiga \& March\~{a} 2004;
Ant\'{o}n \& Browne 2005), and the possible discovery of high
luminosity HBLs is also reported in the Sedentary survey
\citep{giommi05}. In addition, the high-power-high-$\rm \nu_{peak}$
FSRQs were found in the Deep X-ray Radio Blazar Survey (DXRBS),
which is both X-ray and radio-selected, though they do not reach the
extreme $\rm \nu_{peak}$ values of HBLs. From all these discoveries,
it seems that the blazar sequence in its simplest form cannot be
valid (Padovani 2006). By now, what are the main physical parameters
that determines the shift of the synchrotron peak frequency of BL
Lac objects seems still unclear.
%, \cite{jianmin02} suggested that the accretion rate
%influences the shape of the SED in BL Lac objects, and support
%models that couple the jet and the accretion disk.

According to the unified scheme for AGNs, BL Lac objects are
thought to be low-luminosity radio galaxies viewed at relatively
small angles to the line of sight and the relativistic beaming has
an enormous effects on the observed luminosities. Generally, the
intrinsic (debeamed) luminosity emitted from jet in BL Lac objects
can be estimated using the Doppler factor and assuming the jet
geometry. Using the empirical relation of \cite{giova01},
\cite{wu06} estimated the Doppler factor for a sample of 170 BL
Lac objects with the observed radio core observed either with VLA
or MERLIN. In this paper, we will correct the beaming effect in
the multi-band luminosity and the synchrotron peak frequency
directly using the derived Doppler factor of \cite{wu06} for the
same sample. Our aim is mainly to investigate the correlation
between the luminosity at synchrotron peak frequency and the
synchrotron peak frequency after eliminating the beaming effect.
%\cite{capetti02} and \cite{trussoni03}
%have studied the luminosity correlation between BL Lac objects and
%low luminosity radio sources, and indicated that a single
%amplification factor accounts for their difference in luminosity
%in the radio, optical and X-ray band.
The layout of this paper is as follows. The sample is given in Sect.
\ref{sampleselect}. In Sect. \ref{sec:stat}, the various correlation
analysis are presented, which involves the multi-band luminosity,
the synchrotron peak frequency, and the black hole mass as well. The
discussions are shown in Sect. \ref{discussion}, and the results are
summarized in Sect. \ref{conclusion}. Throughout the work, we adopt
the following values for the cosmological parameters: $H_{0}=70 \rm
{~km ~s^ {-1}~Mpc^{-1}}$, $\rm \Omega_{M}=0.3$, and $\rm
\Omega_{\Lambda} = 0.7$, except an otherwise stated. The spectral
index $\alpha$ is defined as $f_{\nu} \propto \nu^{-\alpha}$.

\section{The Sample}

\label{sampleselect} \cite{nieppola2006} presented a large sample
of BL Lac objects from the Mets$\ddot{a}$hovi radio observatory BL
Lac sample consisting of 381 objects selected from the Veron-Cetty
\& Veron BL Lac Catalogue (Veron-Cetty \& Veron 2000), and 17
objects from the literature, of which many sources are from the
well-known BL Lac samples like 1Jy, S4, S5, Einstein Medium
Sensitivity Survey (EMSS), Einstein Slew Survey, and DXRBS. The
authors argued that this sample is supposed to have no selection
criteria (other than declination) in addition to the ones in the
original surveys. From this sample, \cite{wu06} estimated the
Doppler factor for a sample of 170 BL Lac objects using the radio
core emission from either VLA or MERLIN observations. The sample
investigated in this paper is same as that in \cite{wu06}, i.e.
170 BL Lac objects with estimated $\delta$. The Doppler factor was
estimated using the correlation found by \cite{giova01} for radio
galaxies,
\begin{equation}{\rm log~ P_{ci5}=~0.62~log~
P_t+8.41}\end{equation} where P$_{\rm ci5}$ is the intrinsic core 5
GHz radio power derived assuming $\Gamma= 5$ (see \cite{giova01} for
details), and $\rm P_{t}$ is the total radio power at 408 MHz. From
the measured total 408 MHz radio power, we can obtain the intrinsic
core radio power. Thus, the Doppler factor can be got from $\rm
P_{co5}= P_{ci5}\delta^{2+ \alpha}$ (corresponding to a continuous
jet), assuming $\alpha=0$, where $P_{\rm co5}$ is the observed 5 GHz
core luminosity \cite[see][for details]{wu06}.

To construct the SED for all 170 BL Lac objects, we collected the
radio, optical and X-ray data from NASA/IPAC Extragalactic
Database (NED), the Astrophysical Catalogues Support System (CATs)
maintained by the Special Astrophysical Observatory, Russia, and
the catalogue of integral blazar working group
(IBWG)\footnote{http://altamira.asu.cas.cz/iblwg/catalog.php}. The
R band core luminosity were mainly obtained from \cite{urry00},
\cite{Matthew05}, and \cite{Nilsson03}. The X-ray data were
collected from \cite{laure99}, \cite{donato01}, \cite{rec00},
\cite{reich00}, \cite{turriziani07}, and \cite{bauer00}. The
Galactic extinction in optical band were corrected using the value
from NED. Following \cite{nieppola2006}, the SED of each source
were constructed in the $\rm log \it~ \nu ~ - ~ \rm log \it ~\nu
\rm F_{\it \nu}$ representation in the observer's frame, based on
the multi-frequency data. The synchrotron component of the SED was
fitted with a parabolic function $y=Ax^2 + Bx + C$, therefore, the
synchrotron peak frequency in the rest frame was obtained as
$\nu_{\rm peak} =-B(1+z)/2A$, in which $z$ is redshift.

In principle, whether or not to include X-ray data in the fit can be
decided based on the X-ray spectral index fitted from high quality
X-ray spectra. A X-ray spectral index of $\alpha>1$ means a
synchrotron origin of X-ray emission, while $\alpha<1$ indicates an
inverse Compton process origin, which implies that the X-ray data
should be excluded in the fit. Unfortunately, most of sources in our
sample have no well-measured X-ray spectral index, which precludes
us to do so. Tentatively in this work, whether or not to include
X-ray data in the fit was decided solely based on a visually
inspection on the SED of each object. Similar to \cite{wu06}, the
objects were assigned an LBL/IBL/HBL classification according to
$\nu_{\rm peak}$: for LBLs log $\nu_{\rm peak}<$14.5, for IBLs 14.5
$<$ log $\nu_{\rm peak}$ $<$ 16.5, and for HBLs log $\nu_{\rm
peak}>$ 16.5.

We estimated the jet power using the formula derived from
\cite{pun05}:
\begin{equation}
\rm{Q_{jet}=5.7\times10^{44}(1+z)^{1+\alpha}Z^{2}F_{151}} ~~
\rm{erg~s^{-1}}
\end{equation}
\begin{equation}
\rm Z\approx3.31-3.65\times{[(1+z)^{4}-0.203(1+z)^{3}
+0.749(1+z)^{2}+0.444(1+z)+0.205]^{-0.125}}
\end{equation}
where $F_{151}$ is the optically thin flux density from the lobes
measured at 151 MHz in units of janskys, and the value of
$\alpha\approx1$ is a good fiducial value \citep[see][for more
details]{pun05}. We calculated the $F_{151}$ through the
extrapolation from either the extended radio flux at 5 GHz or 1.4
GHz, or the low frequency total radio flux assuming $\alpha$=1.0.

To estimate the black hole mass, we collected the R-band magnitude
of host galaxies for 121 sources, of which the core R-band
magnitude are also available. Following \cite{wu06}, we adopted
the average redshift of the classification (i.e. LBLs, IBLs and
HBLs) for sources without redshift. The sample is listed in Table
\ref{tablesample}: Col. (1) the source IAU name (J2000), Col. (2)
the source alias name, Col. (3) the redshift, Col. (4) the
intrinsic synchrotron peak frequency, Col. (5) the observed
synchrotron peak luminosity, Col. (6) the jet power, Col. (7) the
Doppler factor measured in \cite{wu06}, Cols. (8) - (9) the R-band
magnitude of the host galaxies and the corresponding references,
respectively, Col. (10) the black hole mass.%, Cols. (11) - (12)
%the 2-10 keV luminosity and the corresponding references,
%respectively.

\section{Correlation analysis}
\label{sec:stat} %\cite{nieppola2006} have shown negative
%correlations between the luminosity at 5 GHz, 37 GHz, and 5500\AA~
%and the synchrotron peak frequency $\nu_{\rm peak}$, and a
%positive correlation between the luminosity at 1 keV and $\nu_{\rm
%peak}$. However, they found no significant correlation between the
%luminosity at synchrotron peak frequency and $\nu_{\rm peak}$,
%while there is a weak negative correlation if only the objects
%with log $\nu_{\rm peak} <17$ are considered, and a significant
%positive correlation can be found for the sources with log
%$\nu_{\rm peak}>17$.
It is commonly accepted that the luminosity of blazars is
dominated by the emission from jets, which however is Doppler
boosted. Using the Doppler factor derived in \cite{wu06} and
assuming a same Doppler factor at various wavebands of synchrotron
emission as suggested by \cite{capetti02} and \cite{trussoni03},
we calculate the intrinsic luminosity using $L_{\rm int}=L_{\rm
obs}/\delta^{2+\alpha}$ (corresponding to a continuous jet), in
which $L_{\rm int}$ is the intrinsic luminosity, $L_{\rm obs}$ the
observed luminosity. Since the synchrotron peak frequency
$\nu_{\rm peak}\propto B\delta \gamma_{\rm peak}^{2}$, the
intrinsic one $\nu_{\rm peak}^{'}$ can be obtained as $\nu_{\rm
peak}/\delta$, which is only dependent of $B$ and $\gamma_{\rm
peak}$, where $B$ is the strength of the magnetic field, $\delta$
the Doppler factor, and $\gamma_{\rm peak}$ a characteristic
electron energy that is determined by a competition between
accelerating and cooling processes. With the intrinsic parameters,
we explore various correlations below.
%0, 1.0, 0.7 for the radio(5GHz, 37GHz), optical and X-ray(1 keV) band.

\subsection{Radio and optical core luminosity} \label{radiooptical}
\begin{figure*}
\begin{center}
\includegraphics[width=13cm,angle=0]{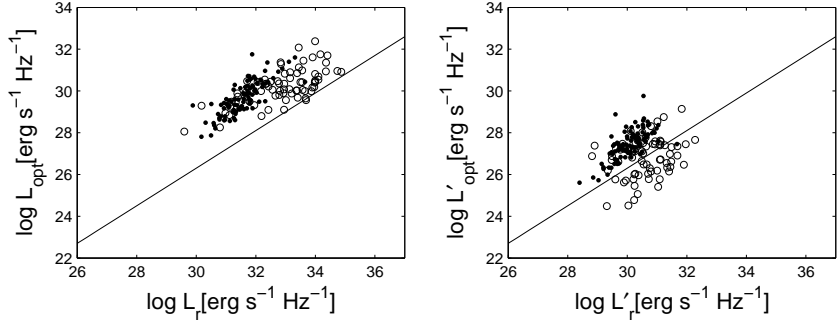}
\caption{The optical vs. radio core luminosity. The solid line is
the best fit of the correlation found by \cite{chiaber01} for FR I
nuclei. The open circles represent LBLs, while the filled circles
are for IBLs and HBLs. In the left panel, the observed luminosities
are shown, while the debeamed luminosities with $\Gamma=5$ for a
mean viewing angle of $60^{\circ}$ are used in the right panel (see
text for details).} \label{raopt}
\end{center}
\end{figure*}

We plot the observed radio and optical core luminosity in the left
panel of Fig. \ref{raopt}, together with the best fit (the solid
line) of the correlation found by \cite{chiaber01} for FR I nuclei.
The objects in the present sample span about five orders of
magnitude in the radio core luminosity, and the optical core
luminosity increase linearly with it over this range. There is a
significant correlation between the radio and optical core
luminosity at $\gg$99.99\% level, which is confirmed by the partial
correlation analysis to exclude the common dependence on the
luminosity distance. This correlation suggests that the emission at
two wavebands is ascribed to the same non-thermal process, just like
in FR I radio galaxies. However, we find that the points are all
above the fitted line of \cite{chiaber01}, and the optical core
luminosity are brighter with about two orders of magnitude than that
of FR I nuclei, at the same radio core luminosity.

We correct the beaming effect for the radio and optical core
luminosity with the estimated Doppler factor. The intrinsic
optical core luminosity are estimated assuming $\alpha_{\rm o}=1$
\citep{falomo94}, while we use $\alpha_{\rm r}=0.0$ in the radio
band.  We also recalculated the luminosities in the hypothesis of
a bulk Lorentz factor of $\Gamma=5$ and a viewing angle of
$60^{\circ}$, which is the average angle expected for the
population of FR I radio galaxies used by \cite{chiaber01} to
derive the correlation. The recalculated luminosities are
presented in the right panel of Fig. \ref{raopt}. It's apparent
that the data points move to the lower radio and optical
luminosity due to the debeaming. Although with some scatter, the
debeamed radio and optical luminosity follow the correlation of
\cite{chiaber01} for FR I nuclei. This supports a common nature of
BL Lac objects and FR I radio galaxy nuclei, which is consistent
with the results of \cite{gir06} from the study of 29 nearby BL
Lac objects.
%\textbf{It
%can also be noted that the HBLs systematically offset from the line,
%while the LBLs cluster around the line. This may indicated that
%there should be some intrinsic differences between the two type of
%BL Lac objects, the spectral index between radio and optical
%$\alpha_{\rm ro}$, radio loudness, or the jet structure.}

%It can also be noted that the HBLs systematically offset from the
%line, while the LBLs cluster around the line. This may be caused by
%the difference of the spectral index between radio and optical
%$\alpha_{\rm ro}$, since LBLs usually have steeper $\alpha_{\rm ro}$
%than that of HBLs \citep{nieppola2006}. Alternatively, a two
%velocity jet (fast spine and slower external layer, with different
%weight in LBL and HBL) can be a plausible explanation for this
%offset \citep{gir06}.
%the Doppler factor got from radio band is
%similar with the Doppler factor in optical band, this result is
%well similar with \cite{gir06} and \cite{capetti02} and
%\cite{trussoni03} that the single amplification factor accounts
%for their difference in luminosity in the radio, optical and X-ray
%band for our BL Lac objects sample.
%The result can be interpret in
%terms of a common synchrotron origin, with little contribution
%from a radiatively efficient accretion disk, as showed in
%\cite{gir06}.

\subsection{Synchrotron peak frequency and luminosity}
%\subsection{The 37GHz luminosity correlation}
%Based on the 37 GHz data in \cite{nieppola07}, in fig \ref{p37}, we
%plot the observed and intrinsic 37 GHz luminosity(Doppler correction
%and redshift correction) versus intrinsic peak frequency , There is
%an anti-correlation between the observed luminosity and $\nu_{peak}$
%with r=-0.38, p=0.90E-03, it is a strong anti-correlation similar
%with the result in \cite{nieppola2006}. We also can see that there
%is no negative correlation for the intrinsic 37 GHz
%luminosity(Doppler correction and redshift correction) versus
%intrinsic peak frequency found by \cite{nieppola2006}. The
%parameters of the correlation are showed in table \ref{table2}.
%There are nearly no correlation. This is also different with the
%correlation of low frequency radio luminosity with synchrotron peak
%frequency found by \cite{wu06}, which is supposed to be the
%intrinsic luminosity.

%\begin{figure*}
%\begin{center}
%\includegraphics[width=18cm,angle=0]{p372170.eps}
%\caption{The 37GHz luminosity selected from \cite{nieppola07}
%against intrinsic synchrotron peak frequency. The left: the original
%data, the right are the corrected data.} \label{p37}
%\end{center}
%\end{figure*}
%\subsection{The optical luminosity correlation}

\begin{figure*}
\begin{center}
\includegraphics[width=18 cm,angle=0,clip]{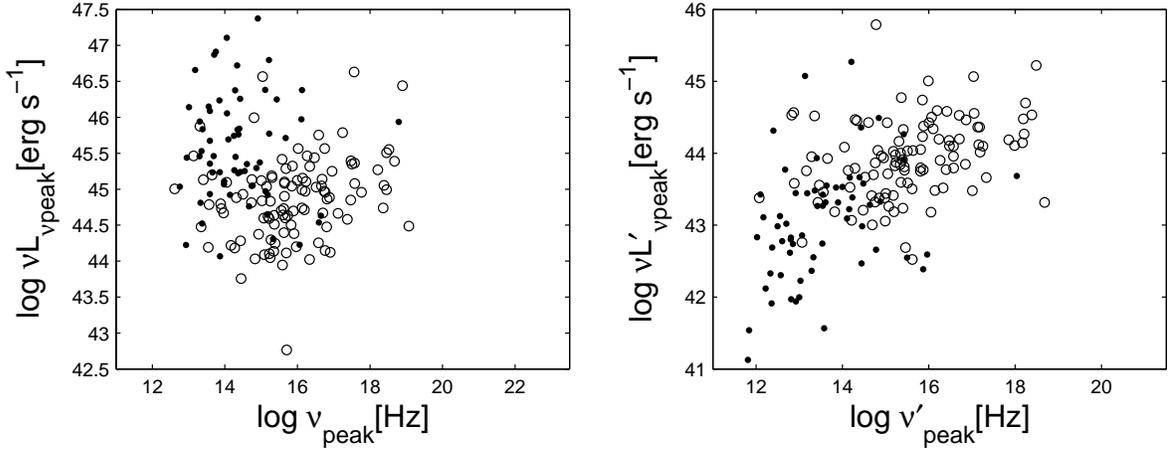}
\caption{The peak luminosity versus synchrotron peak frequency. The
left panel is for observed values, while the intrinsic ones are
shown in the right panel. The open circles represent the sources
with the Doppler factor $\delta < 3.5$ (see text for details).}
\label{pl}
\end{center}
\end{figure*}

In the left panel of Fig. \ref{pl}, we present the relationship
between the observed peak luminosity ${\nu L_{\nu}}_{\rm peak}$ and
synchrotron peak frequency $\nu_{\rm peak}$, while the relation of
the intrinsic parameters ${\nu L_{\nu}}_{\rm peak}^{'}$ and
$\nu_{\rm peak}^{'}$ is shown in the right panel. We find a weak
negative correlation between ${\nu
L_{\nu}}_{\rm peak}$ and $\nu_{\rm peak}$. %, \textbf{which is
%actually similar to that of \cite{nieppola2006} for a larger
%sample if the same range of peak frequency are considered}.
In contrast, after correcting the beaming effect both in the peak
luminosity and synchrotron peak frequency, we find a significant
positive correlation between ${\nu L_{\nu}}_{\rm peak}^{'}$ and
$\nu_{\rm peak}^{'}$ with a Spearman correlation
coefficient of $r = 0.59$ at $\gg$ 99.99\% confidence level. %and this
%result is similar with \cite{valtaoja07}, which have shown some
%unpublished result about the correlation between the synchrotron
%peak luminoisty and synchroton peak frequency after Doppler
%correcting, a slightly positive correlation was present.
It should be noted that this correlation may be caused by the common
dependence of the both parameters on the Doppler factor, as $L_{\rm
int}=L_{\rm obs}/\delta^{2+\alpha}$ ($\alpha=1$ in this case) and
$\nu_{\rm peak}^{'}=\nu_{\rm peak}/\delta$. We therefore use the
partial Spearman correlation method (Macklin 1982) to check this
correlation. Still, a significant correlation with a correlation
coefficient of 0.33 is present at 99.99\% significance level between
${\nu L_{\nu}}_{\rm peak}^{'}$ and $\nu_{\rm peak}^{'}$, independent
of the Doppler factor.
%see the parameters in table \ref{table2}.
To further check this correlation, we also perform a statistic
analysis on the sources in the restricted Doppler factor range
$\delta < $3.5. For this subsample of sources, there is no
correlation between the Doppler factor and $\nu_{\rm peak}^{'}$,
%(\textbf{We also can see that from  Fig 13 of \cite{wu06}}),
while a
significant correlation is still present between ${\nu L_{\nu}}_{\rm
peak}^{'}$ and $\nu_{\rm peak}^{'}$ with a Spearman correlation
coefficient of r=0.37, at over 99\% significance level (see Fig.
\ref{pl}). Therefore, we conclude that the positive correlation
between the intrinsic peak luminosity and intrinsic peak frequency
might be intrinsic, at least for our present sample.

%\begin{figure*}
%\begin{center}
%\includegraphics[width=10cm,angle=0,clip]{peakdop122.eps}
%\caption{The Doppler factor versus the intrinsic synchrotron peak
%frequency. The open circles represent the sources with $\delta <
%3.5$ (see text for details).} \label{pdop}
%\end{center}
%\end{figure*}

%\subsection{The X-ray luminosity correlation}
\label{xray}

The intrinsic luminosity at synchrotron peak frequency can be a
good indicator of the jet emission due to the fact that the
majority of jet synchrotron emission are radiated at the
synchrotron peak frequency. Therefore, the significant positive
correlation between ${\nu L_{\nu}}_{\rm peak}^{'}$ and $\nu_{\rm
peak}^{'}$ implies that the jet emission are tightly related with
the synchrotron peak frequency, which is only dependent of $B$ and
$\gamma_{\rm peak}$.%, \textbf{at lease for this present sample}.

%and this is against the blazar sequence, the original mildly
%negetive correlation might be caused by the Doppler factors for
%different types of BL Lac objects are different.

%\begin{figure}
%\begin{center}
%\includegraphics[width=18cm,angle=0,clip]{popt1121.eps}
%\caption{The left panel: the observed optical luminosity and
%synchrotron peak frequency; the right panel: the intrinsic optical
%luminosity and synchrotron peak frequency. The open circles
%represent the sources with the Doppler factor $\delta < 3.5$ (see
%text for details). } \label{popt}
%\end{center}
%\end{figure}

\begin{figure}
\begin{center}
\includegraphics[width=14.0 cm,angle=0,clip]{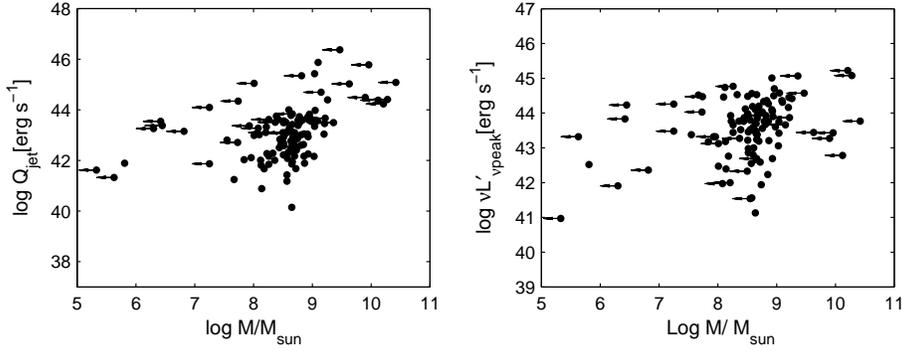}
\caption{Left: the black hole mass versus the jet power. Right:
the black hole mass versus the intrinsic luminosity at synchrotron
peak frequency. The arrows indicate the upper limit of black hole
mass.} \label{mbhlum}
\end{center}
\end{figure}

%\begin{figure}
%\begin{center}
%\includegraphics[width=16 cm,angle=0,clip]{plpedd1121.eps}
%\caption{The relation between the synchrotron peak luminosity in the
%unit of the Eddington luminosity and the synchrotron peak frequency.
%The left panel is for the observed luminosity, while the right panel
%is for the intrinsic luminosity. The open circles are the sources
%with $\delta < 3.5$, and the arrows indicate the lower limit.}
%\label{peaklped}
%\end{center}
%\end{figure}

\subsection{The black hole mass and luminosity} \label{bhandlum}
We estimated the black hole mass using the relation between the
R-band absolute magnitude of host galaxy $M_{\rm R}$ and the black
hole mass $M_{\rm bh}$ proposed by McLure \& Dunlop (2002):
\begin{equation}{ \rm log~ \it M_{\rm bh}=\rm -0.50~\it M_{\rm R}\rm -2.96}
\end{equation}
for 121 sources with available $M_{\rm R}$ from literature. There
are four sources without measured redshift, which are excluded from
the correlation analysis.

Fig. \ref{mbhlum} shows the correlation between the black hole
mass and jet power (left), and intrinsic peak luminosity (right).
The black hole mass of 34 sources are only given as an upper limit
due to the faintness and
the upper limit of R-band magnitude of host galaxies. %And the
%correlation parameters are listed in table \ref{table3},
We used the astronomy Survival Analysis (ASURV) package (Isobe \&
Feigelson 1990; Lavalley et al. 1992) when we
investigate the correlation involving these upper limits.  %We can see that there
%is an weak positive correlation between black hole mass and jet
%power, a little different with the result in \cite{liu06}, which
%have done the correlation for a large sample of Quasars, and
%present a stong positive correlation. And also the correlation
%with the observed and intrinsic radio core luminosity at 5 GHz
%with black hole mass are also weak correaltion, this result are
%similar with the result in \cite{gir06} for a sample of low
%redshift BL Lac objects. The correlation with the total luminosity
%at 37 GHz is stronger, over 99 percent level, for all the 37 GHz
%luminosity, we have used the published data in \cite{nieppola07},
%for sources with more than one observations, we used the averaged
%value.
We find a strong correlation between the intrinsic peak luminosity
${\nu L_{\nu}}_{\rm peak}^{'}$ and the black hole mass $M_{\rm
bh}$ with a correlation coefficient of $r=0.27$ at 99.7\%
confidence level, while there is originally no correlation between
the observed peak luminosity and the black hole mass. The partial
correlation analysis to exclude the common dependence on redshift
still show a significant correlation at 99.9\% significance level.
Moreover, we find a mild correlation between the jet power and the
black hole mass at about 95\% confidence level, which is also
confirmed by the partial correlation analysis. %In addition, a
%strong correlation between the intrinsic 1 keV luminosity and the
%black hole mass is found (at 99.6\% confidence), whereas there is
%only mild correlation in the case of optical band.
These results imply that the sources with larger jet power and/or
jet emission systematically possess heavier black holes.

\section{Discussion}
\label{discussion}

 It can be seen from Fig. \ref{pl} that very high synchrotron peak
frequency exist in several objects with $\rm log \nu_{peak}>17$.
\cite{Ghise99} have suggested the existence of ultra-high-energy
synchrotron peak BL Lacs (UHBLs), whose synchrotron peak frequencies
lies at even higher frequencies than that of conventional HBLs, $\rm
log \nu_{peak}>19$. In our sample, we find eight objects with $\rm
log \nu_{peak}^{'}>18$ (see Table \ref{tablesample}), which thus can
be potential UHBLs. However, as \cite{nieppola2006} pointed out,
using a simple parabolic function in the fitting may produce some
error, especially among HBLs, and the peak frequencies of the most
extreme objects can be exaggerated and cannot be considered as
definite. Although there are no sources with extreme peak frequency
$\rm log \nu_{peak}^{'}>19$ in our sample, we conservatively
re-examine our results after excluding all eight sources with $\rm
log \nu_{peak}^{'}>18$. We find it does not alter our results.

We note that equation (1) used to estimate the Doppler factor is
based on the assumption of a Lorentz factor of five \citep{giova01}.
As \cite{giova01} claimed, the similar correlation can be obtained
for $\Gamma$ in the range 3-10, so generally equation (1) can be
extensively used in this range. However, the simplification of
similar Lorentz factor in a sample may produce errors in the
estimated Doppler factor. As a matter of fact, the superluminal
motion studies \citep{vermeulen94,jorstad01,jorstad05} showed that
some LBLs may have large apparent velocity, $\rm > 10c$ in some
cases, which implies that their Lorentz factor are larger than 10.
We collected the available apparent velocity for 27 sources of our
sample from literatures, and find that the median value is about 3c,
and only three sources with apparent velocity larger than 10c. As BL
Lac objects are normally viewed at small viewing angle, the Lorentz
factor can be roughly constrained if we assume the viewing angle
$\theta=1/\Gamma$. We find that the median value of the constrained
Lorentz factor is right in the range 3-10.
%\cite{kharb07} have
%also showed that LBLs typically show superluminal motions with
%apparent speeds in the range 1 - 5c and the apparent velocity is
%typically less than 2c for the HBLs.
Moreover, utilizing the variability Doppler factor and apparent
velocity, \cite{dopvar} and \cite{valtaoja99} showed that the
averaged lorentz factors of radio selected BL Lac objects are around
5. In addition, \cite{gab00} found no evidence that the typical jet
Lorentz factor exceed $\Gamma \sim5-6$ from the apparent
superluminal speed distribution of 16 northern BL Lac objects. We
thus conclude that the assumption of a Lorentz factor of 5, or a
range of 3-10 is a good approximation although the exceptions do
exist. We find that the sources with large apparent velocity are
mainly LBLs. If these sources do have Lorentz factor larger than
ten, our method may likely underestimate the Doppler factor.
Adopting a real Doppler factor may produce a smaller intrinsic
synchrotron luminosity than ours, which however may strengthen our
results since these sources will be lower down vertically in the LBL
region in Fig. \ref{pl}.

The `blazar sequence', power - $\nu_{\rm peak}$ anti-correlation,
was originally presented by \cite{fossati98} based on one radio- and
one X-ray-selected BL Lacs samples, and one FSRQ sample. These
samples had been assembled in an independent and somewhat different
way, especially so as regarding the selection band, and indeed none
of the individual samples in Fig. 7 of \cite{fossati98} shown the
claimed anti-correlation between power and $\nu_{\rm peak}$, which
was only apparent by combining the three samples \citep{padovani06}.
The blazar sequence was recently tested by \cite{nieppola2008}.
After eliminating the beaming effect on peak frequencies and
luminosities, \cite{nieppola2008} found that the blazar sequence
disappears. Instead, for BL Lac objects separately, they found a
positive correlation between the synchrotron peak frequencies and
luminosities. Intriguingly, this is consistent with our results. The
blazar sequence thus can be an observational phenomenon created by
variable Doppler boosting across the synchrotron peak frequency
range \citep{nieppola2008}.
  By now which really determines the shifting of the synchrotron peak frequency
seems still unclear. \cite{jianmin02} claimed that the shape of SEDs
is related to the accretion rate in BL Lac objects. They found that
HBLs and LBLs follow the different anti-correlation between the peak
frequency and accretion rate, which may indicate differences in the
physical processes associated with the jets. We re-studied their
results by collecting the emission line flux and using the intrinsic
peak frequency for our samples. We found that the overall relation
between the peak frequency and accretion rate is similar to their
results. However, the correlation in HBLs and LBLs separately are
not evident. The difference may be partly because we are using the
intrinsic peak frequency. On the other hand, the number of BL Lac
objects with the detected emission lines is still limited. %%

For a BL Lac sample, the luminosity at any fixed waveband actually
represent different portion of source SED because of the changing
$\nu_{\rm peak}$ in sources. Therefore, the dependence between the
source luminosity and synchrotron peak frequency might be
necessarily checked at the synchrotron peak frequency, at which
the most of jet radiation are emitted. Moreover, it is necessary
to debeam the luminosity, which is Doppler boosted due to the
relatively small viewing angles of jets. The strong positive
correlation between the intrinsic peak luminosity and the
intrinsic synchrotron peak frequency implies that the intrinsic
source luminosity does depend on the location of the synchrotron
peak, although with a large scatter. It seems that the higher
peaked BL Lac objects are intrinsically more powerful than the low
frequency peaked BL Lac objects in the present sample. Since
$\nu_{\rm peak}^{'}\propto \gamma_{\rm peak}^{2}B$, our results
show that the power of jet emission increases with the magnetic
field and electron Lorentz factor. This however is hard to be
explained by the existing models that advocate that the SED of
relativistic jets is strongly
affected by the (external) radiation field \citep[e.g.][]{ghisel98}. %Our results are
%obviously contradict with the `blazar sequence', so the cooling
%process used to explain the blazar seqence can not explain our
%results.
It may be possible that the cooling process is not important in BL
Lac objects. As a matter of fact, the cooling in BL Lac objects is
expected to be smaller than that in Flat Spectrum Radio Quasars
(FSRQs) which have strong broad emission lines. Alternatively, the
efficient injection of the fresh high energy electrons can be
maintained for certain long time, although the cooling process can
not be ignored. In any cases, the theoretical models should be
developed to explain our results. It may also be important if we
could measure the magnetic field for individual sources.

We note that the strong-lined counterparts of HBLs, i.e. FSRQs with
high $\nu_{\rm peak}$ have also been discovered in DXRBS by
\cite{padovani03}. Solely with DXRBS, they found no anti-correlation
between synchrotron peak frequency and radio, BLR and jet powers,
contrary to the predictions of the blazar sequence scenario. Due to
the great similarities between FSRQs and BL Lacs, it might be
crucial to revisit our results with a large and complete sample
including both BL Lacs and FSRQs, which will help us understand the
dependence between the source luminosity and synchrotron peak
frequency, jet formation and physics.
%On
%the other hand, the theoretical model are needed to explain our
%results.
%If the
%positive correlation between the power and $\nu_{\rm peak}$ exist,
%this will support all the observational data showed that HBL
%subclasses makes up a small (10\%) minority of BL Lac objects
%\citep{padovani03}.
\section{Conclusions}
\label{conclusion} \hspace*{2em}By eliminating the beaming effect
with the estimated Doppler factor from the radio data
\citep{wu06}, we have calculated the intrinsic luminosity and the
intrinsic synchrotron peak frequency for a sample of 170 BL Lac
objects. Moreover, we estimated the black hole mass for our
sample. The results can be summarized as follows:

1.~~ After correcting the beaming effects, the radio and optical
core luminosities follow with little scatter the same correlation
found for FR I radio galaxies by \cite{chiaber01}, in support of
the unification of BL Lac objects and FR I galaxies based on
orientation.

2.~~ We find a significant positive correlation between the
intrinsic luminosity at synchrotron peak frequency and the
intrinsic synchrotron peak frequency. The result implies that the
jet emission is tightly related with the synchrotron peak
frequency. %However, this result is in contrast with the `blazar
%sequence'.

3.~~ After correcting the beaming effect, we find a strong
correlation between the black hole mass and the peak luminosity,
and a mild correlation in the case of jet power. The results
indicate that the more powerful jets may have heavier black holes.

%4.~~ The observed radio and X-ray luminosities deviated from the
%fundamental plane of BH activity found by \cite{merloni03} for a
%sample of AGNs and X-ray binaries. However, our sources well follow
%with some scatter the FP relation when we debeam the luminosities.

\begin{acknowledgements}
We thank the anonymous referee for insightful comments and
constructive suggestions. We thank Xinwu Cao for helpful discussions
and suggestions. This work is supported by the NSFC under grants
10373019, 10633010 and 10703009. The National Radio Astronomy
Observatory is operated by Associated Universities, Inc., under
cooperative agreement with the National Science Foundation. MERLIN
is a National Facility operated by the University of Manchester at
Jodrell Bank Observatory on behalf of PPARC. This research made use
of the NASA/ IPAC Extragalactic Database (NED), which is operated by
the Jet Propulsion Laboratory, California Institute of Technology,
under contract with the National Aeronautics and Space
Administration. This work also made use of Astrophysical Catalogues
Support System (CATS) maintained by the Special Astrophysical
Observatory, Russia.

\end{acknowledgements}

\begin{table}
\setlength{\tabcolsep}{0.08in}
 %\tabletypesize{\scriptsize}
 \caption{The sample\label{tablesample}}
 \begin{center}
\begin{tabular}{lccccccccccc}
\hline
 IAU name & Source & $z$ & Log~$\nu_{\rm peak}^{'}$  & Log~$
\nu L_{\rm \nu p} $& Log $Q_{\rm jet}$&$ \delta $& $ m_{\rm h} $&
Refs.& Log  $M_{\rm bh} $\\
\hline(1) & (2)&(3) & (4)  & (5) & (6)& (7) & (8) &(9)& (10)\\\hline
0006$-$063  &NRAO 5            &     0.347  &   12.17   &   45.44   &   44.79  &    5.97  &         ...   &    ...   &    ...    \\
0007+472  &RX J0007.9+4711     &     0.280  &   14.75   &   44.62   &   43.37  &    2.71  &      $>$19.01   &    1 &   7.93   \\
0035+598  &1ES 0033+595        &     0.086  &   17.01   &   44.58   &   41.87  &    2.33  &      $>$17.56   &     2 &   7.25   \\
0040+408  &1ES 0037+405        &     0.271  &   18.17   &   44.99   &   42.79  &    1.92  &         ...   &    ...   &    ...    \\
0050$-$094  &PKS 0048$-$097    &     0.216  &   12.89   &   45.13   &   44.22  &    3.29  &         ...   &    ...   &    ...    \\
0110+418  &NPM1G +41.0022      &     0.096  &   15.23   &   44.05   &   42.70  &    1.06  &       15.28   &   1 &   8.51        \\
0112+227  &S2 0109+22          &     0.473  &   12.92   &   46.23   &   44.17  &    8.50  &         ...   &    ...   &    ...      \\
0115+253  &RXS J0115.7+2519    &     0.350  &   13.47   &   44.80   &   43.59  &    2.59  &       17.91   &      1 &   8.76    \\
0123+343  &1ES 0120+340        &     0.272  &   17.99   &   45.51   &   42.72  &    2.93  &       17.41   &      3 &   8.69    \\
0124+093  &MS 0122.1+0903      &     0.339  &   15.61   &   44.38   &   41.43  &    1.96  &       18.20   &      4 &   8.57    \\
0136+391  &B3 0133+388         &     0.271  &   16.20   &   45.75   &   43.29  &    2.44  &         ...   &    ...   &    ...      \\
0141$-$094  &PKS 0139$-$09     &     0.733  &   13.19   &   46.06   &   45.03  &    7.43  &      $>$18.09   &     4 &   9.63  \\
0148+140  &1ES 0145+138        &     0.125  &   15.86   &   43.94   &   42.01  &    0.54  &       16.67   &      2 &   8.12    \\
0153+712  &8C 0149+710         &     0.022  &   14.72   &   44.03   &   41.86  &    1.29  &       11.68   &       1 &   8.65    \\
0201+005  &MS 0158.5+0019      &     0.299  &   17.85   &   45.27   &   42.32  &    2.30  &       17.84   &       2&   8.59    \\
0208+353  &MS 0205.7+3509      &     0.318  &   15.18   &   44.73   &   41.76  &    2.75  &         ...   &    ...   &    ...       \\
0214+517  &87GB 02109+5130     &     0.049  &   15.51   &   44.12   &   42.33  &    1.50  &       13.56   &       3 &   8.60     \\
0222+430  &3C 66A              &     0.440  &   14.84   &   46.25   &   45.41  &    3.85  &         ...   &    ...   &    ...       \\
0232+202  &1ES 0229+200        &     0.140  &   16.48   &   44.96   &   43.04  &    1.93  &       15.34   &        2&   8.92     \\
0238+166  &AO 0235+164         &     0.940  &   12.68   &   46.87   &   45.09  &   10.78  &      $>$17.18   &     4 &  10.42    \\
0301+346  &MS 0257.9+3429      &     0.245  &   13.29   &   44.78   &   42.34  &    1.90  &       17.18   &      2&   8.68      \\
0314+247  &RXS J0314.0+2445    &     0.054  &   13.36   &   44.47   &   41.25  &    0.97  &       15.65   &     1 &   7.67     \\
0326+024  &2E 0323+0214        &     0.147  &   15.87   &   44.72   &   42.01  &    2.07  &       16.59   &      1 &   8.36      \\
0416+010  &2E 0414+0057        &     0.287  &   16.42   &   45.57   &   43.98  &    2.13  &       16.83   &        2&   9.05      \\
0422+198  &MS 0419.3+1943      &     0.512  &   16.10   &   45.44   &   43.01  &    2.32  &       18.78   &      2&   8.81      \\
0424+006  &PKS 0422+004        &     0.310  &   14.88   &   45.71   &   44.15  &    6.24  &         ...   &    ...   &    ...            \\
0505+042  &RXS J0505.5+0416    &     0.027  &   15.62   &   42.76   &   41.89  &    1.20  &       17.81   &      1 &   5.81     \\
0507+676  &1ES 0502+675        &     0.314  &   18.04   &   45.94   &   42.05  &    5.62  &         ...   &    ...   &    ...     \\
0508+845  &S5 0454+84          &     0.112  &   12.83   &   44.07   &   41.62  &   10.76  &      $>$21.99   &       2&   5.33      \\
0509$-$040  &4U 0506$-$03      &     0.304  &   18.39   &   45.39   &   43.42  &    1.93  &       17.76   &      2 &   8.66        \\
0613+711  &MS 0607.9+7108      &     0.267  &   14.48   &   44.84   &   41.89  &    3.49  &       17.00   &     2 &   8.87        \\
0625+446  &87GB 06216+4441     &     0.473  &   13.36   &   45.69   &   44.53  &    5.46  &         ...   &    ...   &    ...            \\
0650+250  &1ES 0647+250        &     0.203  &   15.51   &   45.57   &   42.71  &    3.24  &      $>$18.61   &      2 &   7.73      \\
0654+427  &B3 0651+428         &     0.126  &   14.98   &   44.51   &   42.93  &    2.38  &       15.59   &      1 &   8.67        \\
0656+426  &NPM1G +42.0131      &     0.059  &   15.41   &   44.14   &   43.73  &    0.86  &       13.63   &     1 &   8.78        \\
0710+591  &EXO 0706.1+5913     &     0.125  &   16.85   &   44.66   &   42.85  &    1.87  &       15.57   &      1 &   8.67        \\
0721+713  &S5 0716+714         &     0.300  &   14.30   &   45.99   &   44.35  &    3.23  &      $>$19.55   &     2 &   7.74      \\
0738+177  &PKS 0735+17         &     0.424  &   12.82   &   46.38   &   43.28  &   29.41  &      $>$19.75   &     2 &   8.08      \\
0744+745  &MS 0737.9+7441      &     0.315  &   14.52   &   45.12   &   42.60  &    3.05  &       17.55   &      2&   8.81        \\
0753+538  &S4 0749+54          &     0.200  &   12.36   &   44.81   &   43.26  &    9.26  &      $>$21.44   &     2 &   6.30      \\
0757+099  &PKS 0754+100        &     0.266  &   13.00   &   45.74   &   43.11  &   17.72  &      $>$18.31   &      2&   8.21      \\
0806+595  &SBS 0802+596        &     0.300  &   15.31   &   45.28   &   43.04  &    2.36  &       16.63   &     1 &   9.20        \\
0809+523  &1ES 0806+524        &     0.137  &   16.14   &   45.04   &   42.69  &    3.32  &       15.83   &     3&   8.65        \\
0818+423  &OJ 425              &     0.530  &   12.55   &   45.53   &   45.05  &    6.33  &      $>$20.47   &    2 &   8.01    \\
0823+223  &4C 22.21            &     0.951  &   12.87   &   45.88   &   46.38  &    2.73  &      $>$19.10   &    4 &   9.47    \\
0825+031  &PKS 0823+033        &     0.506  &   11.84   &   45.94   &   43.08  &   29.42  &      $>$19.28   &     2 &   8.55    \\
0831+044  &PKS 0829+046        &     0.174  &   13.07   &   45.24   &   43.56  &    6.21  &       16.66   &      2 &   8.52      \\
0831+087  &1H 0827+089         &     0.941  &   14.00   &   45.35   &   44.95  &    4.04  &         ...   &    ...   &    ...    \\
0832+492  &OJ 448              &     0.548  &   12.38   &   45.46   &   43.84  &    8.39  &      $>$19.26   &    2  &   8.66      \\
0854+441  &US 1889             &     0.382  &   15.04   &   45.19   &   43.90  &    1.79  &         ...   &    ...   &    ...           \\
0854+201  &OJ 287              &     0.306  &   12.33   &   46.09   &   43.10  &   17.87  &      $>$18.08   &    2 &   8.50     \\
0915+295  &B2 0912+29          &     0.302  &   15.39   &   45.32   &   44.01  &    2.96  &         ...   &    ...   &    ...           \\
0916+526  &RXS J0916.8+5238    &     0.190  &   15.36   &   44.40   &   43.55  &    1.85  &       15.95   &      1&   8.98        \\
0929+502  &RXS J0929.2+5013    &     0.370  &   13.33   &   45.45   &   43.86  &    9.23  &         ...   &    ...   &    ...         \\
0930+498  &1ES 0927+500        &     0.188  &   17.33   &   44.96   &   41.85  &    2.70  &       17.36   &     2 &   8.26        \\
0930+350  &B2 0927+35          &     0.302  &   12.81   &   44.52   &   44.29  &    3.68  &         ...   &    ...   &    ...          \\
0952+656  &RGB J0952+656       &     0.302  &   14.96   &   44.83   &   43.24  &    1.99  &         ...   &    ...   &    ...          \\
0954+492  &MS 0950.9+4929      &     0.380  &   17.25   &   45.08   &   42.06  &    2.12  &         ...   &    ...   &    ...          \\
0958+655  &S4 0954+65          &     0.368  &   13.54   &   45.76   &   43.51  &    6.79  &      $>$18.81   &    2  &   8.37        \\
\hline
\end{tabular}
\end{center}
\end{table}
\addtocounter{table}{-1}
\begin{table}
\setlength{\tabcolsep}{0.08in}
\caption{Continued\dots\label{tablesample}}
 \begin{center}
\begin{tabular}{lccccccccccc}
\hline
 IAU name & Source & $z$ & Log~$\nu_{\rm peak}^{'}$  & Log~$
\nu L_{\rm \nu p} $& Log $Q_{\rm jet}$&$ \delta $& $ m_{\rm h} $&
Refs.& Log  $M_{\rm bh} $\\
\hline(1) & (2)&(3) & (4)  & (5) & (6)& (7) & (8) &(9)& (10)\\\hline
1012+424  &RXS J1012.7+4229    &     0.364  &     18.24   &   45.55  &   43.69  &    1.92  &       17.61   &     1 &   8.96        \\
1015+494  &GB 1011+496         &     0.200  &     16.27   &   45.14  &   43.61  &    2.59  &       16.15   &     3 &   8.94        \\
1031+508  &1ES 1028+511        &     0.360  &     16.71   &   45.79  &   43.10  &    3.39  &       18.10   &     3 &   8.70        \\
1037+571  &RXS J1037.7+5711    &     0.302  &     14.40   &   45.37  &   43.71  &    3.70  &         ...   &    ...   &    ...   \\
1047+546  &1ES 1044+549        &     0.473  &     12.83   &   45.46  &   42.96  &    2.05  &       19.31   &       2&   8.44        \\
1053+494  &MS 1050.7+4946      &     0.140  &     14.99   &   44.25  &   42.16  &    2.49  &       15.12   &       1&   9.03        \\
1104+382  &MRK 421             &     0.030  &     16.40   &   44.67  &   42.27  &    1.99  &       13.22   &      2 &   8.23        \\
1109+241  &1ES 1106+244        &     0.460  &     16.57   &   45.25  &   43.78  &    2.42  &       18.86   &      2&   8.64       \\
1120+422  &EXO 1118.0+4228     &     0.124  &     16.33   &   44.26  &   42.38  &    1.76  &         ...   &    ...   &    ...            \\
1136+701  &MRK 180             &     0.045  &     16.86   &   44.15  &   43.02  &    0.78  &       14.37   &      2 &   8.10       \\
1136+676  &RXS J1136.5+6737    &     0.134  &     15.40   &   44.50  &   41.95  &    2.64  &       15.76   &      3&   8.66       \\
1149+246  &EXO 1449.9+2455     &     0.402  &     17.13   &   45.39  &   43.60  &    2.20  &       18.12   &    1 &   8.83      \\
1150+242  &B2 1147+245         &     0.200  &     13.29   &   44.92  &   43.15  &    7.11  &      $>$20.40   &    2 &   6.82    \\
1151+589  &RXS J1151.4+5859    &     0.302  &     15.43   &   45.04  &   43.79  &    2.58  &         ...   &    ...   &    ...            \\
1209+413  &B3 1206+416         &     0.302  &     13.54   &   45.22  &   43.50  &    6.70  &         ...   &    ...   &    ...            \\
1215+075  &1ES 1212+078        &     0.136  &     14.68   &   44.60  &   43.13  &    2.49  &       15.81   &     2 &   8.65         \\
1217+301  &B2 1215+30          &     0.130  &     14.69   &   44.59  &   43.28  &    3.39  &       15.78   &      2 &   8.61          \\
1220+345  &GB2 1217+348        &     0.643  &     13.63   &   45.84  &   44.42  &    5.80  &         ...   &    ...   &    ...        \\
1221+301  &PG 1218+304         &     0.182  &     17.16   &   45.36  &   42.65  &    2.60  &       16.86   &     2 &   8.47          \\
1221+282  &ON 231              &     0.102  &     14.45   &   44.65  &   42.99  &    5.32  &       16.43   &      2 &   8.01          \\
1223+806  &S5 1221+80          &     0.473  &     14.23   &   45.29  &   44.77  &    4.32  &         ...   &    ...   &    ...        \\
1224+246  &MS 1221.8+2452      &     0.218  &     13.43   &   44.73  &   42.11  &    2.96  &       18.32   &     2 &   7.96          \\
1230+253  &RXS J1230.2+2517    &     0.135  &     14.11   &   44.76  &   43.08  &    3.60  &         ...   &    ...   &    ...             \\
1231+642  &MS 1229.2+6430      &     0.163  &     15.19   &   44.60  &   42.18  &    2.96  &       16.15   &    2  &   8.70           \\
1237+629  &MS 1235.4+6315      &     0.297  &     15.07   &   44.70  &   42.24  &    3.02  &         ...   &    ...   &    ...             \\
1241+066  &1ES 1239+069        &     0.150  &     18.68   &   44.49  &   41.33  &    2.45  &      $>$22.08   &    2  &   5.63        \\
1248+583  &PG 1246+586         &     0.847  &     14.17   &   46.80  &   44.70  &   11.10  &      $>$19.43   &    2 &   9.15        \\
1253+530  &S4 1250+53          &     0.302  &     14.64   &   44.92  &   44.26  &    3.52  &         ...   &    ...   &    ...            \\
1257+242  &1ES 1255+244        &     0.141  &     16.70   &   44.86  &   42.29  &    1.30  &       16.53   &     2 &   8.33         \\
1310+325  &AUCVn               &     0.996  &     12.61   &   47.10  &   44.38  &   27.73  &      $>$17.93   &     4 &  10.12       \\
1322+081  &1ES 1320+084N       &     1.500  &     13.14   &   46.91  &   44.41  &    4.09  &      $>$18.71   &     2 &  10.28      \\
1341+399  &RXS J1341.0+3959    &     0.163  &     18.21   &   44.74  &   43.05  &    1.44  &       15.82   &    1 &   8.86         \\
1402+159  &MC 1400+162         &     0.244  &     16.24   &   45.03  &   44.40  &    1.90  &         ...   &    ...   &    ...             \\
1404+040  &MS 1402.3+0416      &     0.344  &     15.36   &   45.42  &   43.62  &    1.64  &      $>$18.87   &     2 &   8.26      \\
1409+596  &MS 1407.9+5954      &     0.496  &     15.81   &   44.97  &   43.35  &    2.55  &       18.25   &     2 &   9.04        \\
1415+485  &RGB J1415+485       &     0.496  &     14.16   &   45.05  &   43.50  &    4.06  &       19.30   &      1 &   8.51        \\
1415+133  &PKS 1413+135        &     0.247  &     12.08   &   45.00  &   44.52  &    3.47  &         ...   &    ...   &    ...             \\
1417+257  &2E 1415+2557        &     0.237  &     17.17   &   45.35  &   43.57  &    2.13  &       16.28   &      3 &   9.09         \\
1419+543  &OQ 530              &     0.152  &     12.93   &   45.11  &   42.53  &   11.38  &       15.90   &      2 &   8.74         \\
1427+238  &PKS 1424+240        &     0.160  &     15.22   &   45.10  &   43.55  &    2.66  &      $>$20.67   &     2 &   6.42       \\
1427+541  &RGB J1427+541       &     0.106  &     14.13   &   44.18  &   42.17  &    1.38  &       14.81   &     1 &   8.86         \\
1428+426  &H 1426+428          &     0.129  &     18.19   &   45.06  &   42.22  &    1.56  &       15.97   &     2 &   8.51         \\
1439+395  &PG 1437+398         &     0.344  &     16.07   &   45.32  &   43.83  &    1.87  &       17.48   &     2 &   8.95        \\
1442+120  &1ES 1440+122        &     0.163  &     15.81   &   44.74  &   43.10  &    2.06  &       16.45   &     2 &   8.55         \\
1444+636  &MS 1443.5+6349      &     0.298  &     16.58   &   44.48  &   42.17  &    1.69  &         ...   &    ...   &    ...            \\
1448+361  &RXS J1448.0+3608    &     0.271  &     14.90   &   45.16  &   43.31  &    2.51  &         ...   &    ...   &    ...            \\
1458+373  &B3 1456+375         &     0.333  &     12.85   &   44.93  &   43.76  &    5.39  &         ...   &    ...   &    ...            \\
1501+226  &MS 1458.8+2249      &     0.235  &     14.46   &   44.97  &   42.59  &    4.59  &       17.40   &      2 &   8.52           \\
1509+559  &SBS 1508+561        &     2.025  &     14.21   &   47.38  &   44.61  &    5.03  &         ...   &    ...   &    ...         \\
1516+293  &RXS J1516.7+2918    &     0.130  &     14.82   &   44.37  &   42.87  &    1.29  &       15.52   &      1 &   8.75       \\
1517+654  &1H 1515+660         &     0.702  &     17.04   &   46.63  &   43.50  &    3.32  &      $>$18.51   &     2 &   9.36     \\
1532+302  &RXS J1532.0+3016    &     0.064  &     15.01   &   44.10  &   42.17  &    1.66  &       14.86   &      1 &   8.26       \\
1533+342  &RXS J1533.4+3416    &     0.810  &     15.42   &   45.97  &   43.55  &    4.86  &         ...   &    ...   &    ...    \\
1534+372  &RGB J1534+372       &     0.143  &     13.82   &   44.22  &   41.81  &    2.21  &       16.97   &   1 &   8.13        \\
1535+533  &1ES 1533+535        &     0.890  &     18.49   &   46.44  &   44.24  &    2.55  &      $>$17.44   &     4 &  10.21      \\
1536+016  &MS 1534.2+0148      &     0.312  &     17.19   &   44.85  &   43.39  &    1.89  &       17.62   &      2 &   8.76        \\
1540+819  &1ES 1544+820        &     0.271  &     15.63   &   45.12  &   43.21  &    2.30  &      $>$19.14   &     2 &   7.82      \\
1540+147  &4C 14.6             &     0.605  &     13.54   &   45.83  &   45.43  &    6.34  &       18.76   &    1 &   9.04        \\
1542+614  &RXS J1542.9+6129    &     0.302  &     13.90   &   45.25  &   43.06  &    4.42  &         ...   &    ...   &    ...      \\
\hline
\end{tabular}
\end{center}
\end{table}
\addtocounter{table}{-1}
\begin{table} \setlength{\tabcolsep}{0.08in}
\caption{Continued\dots\label{tablesample}}
 \begin{center}
\begin{tabular}{lccccccccccc}
\hline
 IAU name & Source & $z$ & Log~$\nu_{\rm peak}^{'}$  & Log~$
\nu L_{\rm \nu p} $& Log $Q_{\rm jet}$&$ \delta $& $ m_{\rm h} $&
Refs.& Log  $M_{\rm bh} $\\
\hline(1) & (2)&(3) & (4)  & (5) & (6)& (7) & (8) &(9)& (10)\\\hline
1554+201  &MS 1552.1+2020      &     0.222  &    15.81   &   44.99  &   42.94  &    2.06  &       16.44   &     1 &   8.93        \\
1555+111  &PG 1553+11          &     0.360  &    15.42   &   46.38  &   44.10  &    5.07  &      $>$20.99   &     2 &   7.25     \\
1602+308  &RXS J1602.2+3050    &     0.302  &    15.34   &   44.64  &   43.64  &    1.85  &         ...   &    ...   &    ...          \\
1626+352  &RXS J1626.4+3513    &     0.497  &    15.24   &   45.08  &   43.67  &    2.53  &      $>$17.89   &      1 &   9.22       \\
1644+457  &RXS J1644.2+4546    &     0.225  &    15.44   &   44.63  &   43.52  &    1.95  &       16.74   &     1 &   8.79         \\
1652+403  &RGB J1652+403       &     0.240  &    14.05   &   44.88  &   42.89  &    1.85  &         ...   &    ...   &    ...           \\
1653+397  &MRK 501             &     0.034  &    15.96   &   44.63  &   41.68  &    4.79  &       12.51   &     1 &   8.72        \\
1704+716  &RXS J1704.8+7138    &     0.350  &    14.76   &   45.04  &   42.84  &    2.45  &       18.52   &   1 &   8.45        \\
1719+177  &PKS 1717+177        &     0.137  &    12.22   &   44.22  &   43.05  &    5.03  &         ...   &    ...   &    ...           \\
1724+400  &B2 1722+40          &     1.049  &    14.44   &   46.38  &   45.66  &    4.72  &         ...   &    ...   &    ...           \\
1725+118  &H 1722+119          &     0.170  &    15.86   &   45.46  &   43.38  &    2.57  &     $ >$20.75   &      2 &   6.45      \\
1728+502  &IZw187              &     0.055  &    16.05   &   44.02  &   42.03  &    1.91  &       15.36   &      2 &   7.84        \\
1739+476  &OT 465              &     0.473  &    12.02   &   45.03  &   44.93  &    5.42  &     $ >$19.75   &       2 &   8.22     \\
1742+597  &RGBJ 1742+597       &     0.400  &    13.85   &   45.24  &   43.51  &    3.73  &       18.67   &     1 &   8.55         \\
1743+195  &NPM1G +19.0510      &     0.084  &    15.46   &   44.20  &   42.26  &    3.19  &       14.14   &     1 &   8.93         \\
1745+398  &B3 1743+398B        &     0.267  &    16.01   &   45.11  &   44.40  &    1.69  &       16.22   &      1 &   9.26         \\
1747+469  &B3 1746+470         &     1.484  &    12.50   &   46.15  &   45.10  &   11.33  &         ...   &    ...   &    ...          \\
1748+700  &S4 1749+70          &     0.770  &    13.42   &   46.26  &   44.49  &    9.94  &     $ >$17.68   &     2 &   9.90    \\
1749+433  &B3 1747+433         &     0.473  &    13.61   &   45.26  &   44.48  &    4.46  &         ...   &    ...   &    ...             \\
1750+470  &RXS J1750.0+4700    &     0.160  &    17.05   &   44.12  &   43.14  &    0.72  &       16.33   &       1 &   8.58        \\
1751+096  &PKS 1749+096        &     0.322  &    11.81   &   45.83  &   42.78  &   37.09  &       17.93   &     2 &   8.64        \\
1756+553  &RXS J1756.2+5522    &     0.271  &    16.47   &   44.97  &   42.50  &    1.84  &         ...   &    ...   &    ...            \\
1757+705  &MS 1757.7+7034      &     0.407  &    13.17   &   45.20  &   42.51  &    3.02  &       18.93   &     2 &   8.44       \\
1800+784  &S5 1803+784         &     0.680  &    13.41   &   46.72  &   45.35  &    8.51  &      $>$19.51   &     2 &   8.82      \\
1806+698  &3C 371              &     0.051  &    14.18   &   44.28  &   43.75  &    1.79  &       14.07   &     1 &   8.39       \\
1808+468  &RGB J1808+468       &     0.450  &    14.29   &   45.14  &   43.63  &    2.80  &       19.22   &      1 &   8.42        \\
1811+442  &RGB J1811+442       &     0.350  &    15.99   &   44.70  &   43.70  &    0.79  &       17.58   &      1&   8.92        \\
1813+317  &B2 1811+31          &     0.117  &    14.85   &   44.09  &   42.81  &    1.73  &       17.66   &      1&   7.55       \\
1824+568  &4C 56.27            &     0.664  &    12.40   &   46.14  &   45.88  &    4.07  &       18.88   &      2&   9.10        \\
1829+540  &RXS J1829.4+5402    &     0.302  &    14.61   &   44.81  &   43.51  &    1.34  &         ...   &    ...   &    ...      \\
1838+480  &RXS J1838.7+4802    &     0.300  &    13.65   &   45.09  &   43.33  &    2.45  &       18.61   &     1 &   8.22        \\
1841+591  &RGB J1841+591       &     0.530  &    14.34   &   44.93  &   43.46  &    1.44  &       18.08   &      1&   9.21        \\
1853+672  &1ES 1853+671        &     0.212  &    13.07   &   44.19  &   41.67  &    3.00  &       17.83   &    2 &   8.18         \\
1927+612  &S4 1926+61          &     0.473  &    12.70   &   45.67  &   44.43  &    7.66  &         ...   &    ...   &    ...      \\
1959+651  &1ES 1959+650        &     0.047  &    15.87   &   44.54  &   40.89  &    5.20  &       14.40   &    2 &   8.14        \\
2005+778  &S5 2007+77          &     0.342  &    12.80   &   45.46  &   44.00  &    7.70  &       18.16   &      2 &   8.60         \\
2009+724  &S5 2010+72          &     0.473  &    12.79   &   45.23  &   44.80  &    7.45  &         ...   &    ...   &    ...           \\
2022+761  &S5 2023+76          &     0.473  &    14.49   &   45.77  &   44.43  &    5.39  &         ...   &    ...   &    ...           \\
2039+523  &1ES 2037+521        &     0.053  &    14.78   &   44.31  &   40.15  &    3.55  &       13.65   &     2 &   8.65          \\
2134$-$018  &PKS 2131$-$021    &     1.285  &    12.10   &   46.66  &   45.78  &   11.93  &      $>$18.94   &       2 &   9.96        \\
2145+073  &MS 2143.4+0704      &     0.237  &    13.56   &   44.67  &   42.99  &    2.53  &       17.42   &      2 &   8.52          \\
2152+175  &PKS 2149+17         &     0.473  &    12.57   &   45.35  &   43.88  &   10.39  &      $>$20.63   &     2 &   7.78      \\
2202+422  &BL LAC              &     0.070  &    13.58   &   45.05  &   42.08  &   14.47  &       14.42   &      2 &   8.58        \\
2250+384  &B3 2247+381         &     0.119  &    15.05   &   44.64  &   42.51  &    2.03  &       15.51   &      1&   8.64        \\
2257+077  &PKS 2254+074        &     0.190  &    13.03   &   45.07  &   42.63  &    8.85  &       16.22   &     2&   8.85        \\
2319+161  &Q J2319+161         &     0.302  &    15.34   &   44.90  &   43.41  &    2.00  &         ...   &    ...   &    ...        \\
2322+346  &TEX 2320+343        &     0.098  &    15.18   &   44.30  &   42.41  &    1.23  &       14.88   &       1&   8.73         \\
2323+421  &1ES 2321+419        &     0.059  &    14.21   &   43.76  &   41.39  &    1.70  &         ...   &    ...   &    ...           \\
2329+177  &1ES 2326+174        &     0.213  &    16.57   &   44.89  &   42.92  &    2.04  &       17.17   &      2 &   8.51          \\
2339+055  &MS 2336.5+0517      &     0.740  &    14.78   &   46.57  &   44.01  &    1.82  &         ...   &    ...   &    ...             \\
2347+517  &1ES 2344+514        &     0.044  &    15.50   &   44.23  &   41.18  &    3.63  &       13.39   &      2 &   8.57       \\
2350+196  &MS 2347.4+1924      &     0.515  &    15.88   &   45.15  &   43.07  &    1.81  &         ...   &    ...   &    ...       \\
\hline

\end{tabular}
\end{center}

{  Notes: Col. 1: the source IAU name (J2000). Col. 2: the source
alias name. Col. 3: the redshift. Col. 4: the intrinsic synchrotron
peak frequency. Col. 5 the observed peak luminosity in units of $\rm
erg~ s^{-1}$. Col. 6 the jet power in units of $\rm erg~ s^{-1}$.
Col. 7 the Doppler factor. Col. 8 the R magitude of host galaxies.
`$>$' indicates the lower limit. Col. 9 the references of $m_{\rm
h}$:
%1 3.0, 2 1.0, 3, 4.0, 4, 2.0
1, \cite{Nilsson03}.  2, \cite{urry00}. 3, \cite{Nilsson07}. 4,
\cite{Matthew05}.  Col. 10
the black hole mass in units of solar mass.} %Col. 11 the 2-10 keV
%luminosity in units of $\rm erg~
% s^{-1}$. Col. 12 the references of 2-10 keV luminosity.}
%% You can append references to a table using the \tablerefs command.
\end{table}


\begin{thebibliography}{99}
%% you can type \apj for ApJ, \aap for A&A, \apss for Ap&SS, etc. Please consult
%% the macro chjaa.cls. You can also find them in aasguide.tex (AASTeX for ApJ, AJ, PASP)
%% Please follow the format of ChJAA's reference list
\bibitem[RBSC-NVSS (2000) ]{bauer00}
Bauer F.E., Condon J.J., Thuan T.X., Broderick J.J. 2000, ApJS, 129,
547

\bibitem[Capetti et al.(2002)]{capetti02}
Capetti, A., Celotti, A., Chiaberge, M., et al. 2002, \aap, 383, 104

\bibitem[Chiaberge et al.(1999)]{chiaber01}
Chiaberge, M., Capetti, A., Celotti, A., et al. 1999, \aap, 349, 77

\bibitem[Donato et al. (2001)]{donato01}
Donato, D., Ghisellini, G., Tagliaferri, G., Fossati, G. 2001, \aap,
375, 739

\bibitem[L{\"a}hteenm{\"a}ki \& Valtaoja (1999) ]{dopvar}
L{\"a}hteenm{\"a}ki, A.\ \& Valtaoja, E.\ 1999, \apj, 521, 493

\bibitem[Falomo et al. (1994)]{falomo94}
Falomo, R., Scarpa, R., \& Bersanelli, M. 1994, \apjs, 93, 125

\bibitem[Fossati et al.(1998)]{fossati98}
Fossati, G., Maraschi, L., Celotti, A., Comastri, A., \& Ghisellini,
G. 1998, \mnras, 299, 433

\bibitem[Gabuzda et al. (2000)]{gab00}
Gabuzda, D. C., Pushkarev, A. B., \& Cawthorne, T. V. 2000, \mnras, 319, 1109

\bibitem[Ghisellini et al.(1999)]{Ghise99}
Ghisellini, G. 1999, ApL\&C, 39, 17

\bibitem[Giommi et al.(1995)]{Giommi95}
Giommi, P., Ansari, S. G., \& Micol, A. 1995, \aaps, 109, 267

\bibitem[Giommi et al.(2005)]{giommi05}
Giommi, P., Piranomonte, S., Perri, M., \& Padovani, P. 2005, \aap,
434, 385

\bibitem[Giovannini et al.(2001)]{giova01}
Giovannini, G., Cotton, W. D., Feretti, L., Lara, L., \& Venturi, T.
2001, \apj, 552, 508

\bibitem[Giroletti et al.(2006)]{gir06}
Giroletti, M., Giovannini, G., Taylor, G. B., Falomo, R. 2006, \apj,
646, 801

\bibitem[Ghisellini et al.(1998)]{ghisel98}
Ghisellini, G., Celotti, A., Fossati, G., Maraschi, L. \& Comastri,
A. 1998, \mnras, 301, 451

\bibitem[Jorstad et al. (2001)]{jorstad01}
Jorstad, S.~G., Marscher, A.~P., Mattox, J.~R., et al. 2001, \apj,
556, 738

\bibitem[Jorstad et al. (2005)]{jorstad05}
Jorstad, S.~G., Marscher, A.~P., Lister, M.~L., Stirling, A. M.,
Cawthorne, T. V., et al. 2005, \apj, 130, 1418

\bibitem[Laurent-Muehleisen et al.(1999)]{laure99}
Laurent-Muehleisen, S. A., Kollgaard, R. I., Feigelson, E. D.,
Brinkmann, W., Siebert, J. 1999, \apj, 525, 127

\bibitem[Merloni et al. (2003)]{merloni03}
Merloni, A., Heinz, S., \& di Matteo, T. 2003, \mnras, 345, 1057

\bibitem[Nieppola et al.(2006)]{nieppola2006}
Nieppola, E., Tornikoski, M., \& Valtaoja, E. 2006, \aap, 445, 441

\bibitem[Nieppola et al.(2008)]{nieppola2008}
Nieppola, E., Valtaoja, E., Tornikoski, M., Hovatta, T., Kotiranta,
M. 2008, arXiv0803.0654N

\bibitem[Nilsson et al.(2003)]{Nilsson03}
Nilsson, K., Pursimo, T., Heidt, J., Takalo, L. O.,
Sillanp$\ddot{\rm a}\ddot{\rm a}$, A., et al. 2003, \aap, 400, 95

\bibitem[Nilsson et al.(2007)]{Nilsson07}
Nilsson, K., Pasanen, M., Takalo, L. O., Lindfors, E., Berdyugin, A.
et al. 2007, arXiv:0709.2533

\bibitem[O'Dowd \& Urry (2005)]{Matthew05}
O'Dowd, M., Urry, C. M., 2005, \apj, 627, 970

\bibitem[Padovani \& Giommi(1995)]{padovani95}
Padovani, P., \& Giommi, P. 1995, \apj, 446, 547

\bibitem[Padovani et al.(2003)]{padovani03}
Padovani, P., Perlman, E. S., Landt, H., Giommi, P., \& Perri, M.
2003, \apj, 588, 128

\bibitem[Padovani (2007)]{padovani06}
Padovani, P. 2007, Ap\&SS, 309, 63

\bibitem[Punsly (2005)]{pun05}
Punsly, B. 2005, \apj, 623, L9

\bibitem[Rector \& Stocke (2000)]{rec00}
Rector T. \& Stocke J. T. 2003, \aj, 120, 1626

\bibitem[Reich et al.(2000)]{reich00}
Reich W., Fuerst E., Reich P., Kothes R., Brinkmann W., 2000, \aap,
363, 141

\bibitem[Trussoni et al.(2003)]{trussoni03}
Trussoni, E., Capetti, A., Celotti, A., Chiaberge, M., Feretti, L.,
2003, \aap, 403, 889

\bibitem[Turriziani et al.(2007)]{turriziani07}
Turriziani, S., Cavazzuti, E., Giommi, P. 2007, arXiv:  0705.1498

\bibitem[Urry et al.(2000)]{urry00}
Urry, C. M., Scarpa, R., O¡¯Dowd, M., Falomo, R., Pesce, J. E., \&
Treves, A. 2000, ApJ, 532, 816

\bibitem[Urry \& Padovani (1995)]{urry95}
Urry, C. M., \& Padovani, P. 1995, \pasp, 107 , 803

\bibitem[Valtaoja et al. (1999)]{valtaoja99}
Valtaoja, E., L$\ddot{a}$hteenm$\ddot{a}$ki, A.,
Ter$\ddot{a}$sranta, H., Lainela, M., 1999, \iaucirc, 159, 1999

\bibitem[Vermeulen \& Cohen (1994)]{vermeulen94}
Vermeulen, R. C., \& Cohen, M. H., 1994, \apj, 430, 467

\bibitem[Veron-Cetty \& Veron (2000)]{veron00}
Veron-Cetty, M. P. \& Veron, P. 2000, ESOSR, 19, 1

\bibitem[Wang et al. (2002)]{jianmin02}
Wang, J.-M., Staubert, R., \& Ho, L. C. 2002, \apj, 579, 554

\bibitem[Wu et al.(2007)]{wu06}
Wu, Z. Z., Jiang, D., R., Gu, M. F., Liu, Y., 2007, \aap. 466, 63

\end{thebibliography}
\end{document}